\documentclass[conference]{IEEEtran}
\IEEEoverridecommandlockouts

\usepackage{cite}
\usepackage{amsmath,amssymb,amsfonts}
\usepackage{algorithmic}
\usepackage{graphicx}
\usepackage{textcomp}
\usepackage{xcolor}
\usepackage{booktabs}
\usepackage{caption}
\usepackage{float} 
\usepackage{subcaption}

\usepackage[lined,boxed,commentsnumbered,ruled,linesnumbered]{algorithm2e}

\def\BibTeX{{\rm B\kern-.05em{\sc i\kern-.025em b}\kern-.08em
    T\kern-.1667em\lower.7ex\hbox{E}\kern-.125emX}}

\DeclareRobustCommand*{\IEEEauthorrefmark}[1]{%
    \raisebox{0pt}[0pt][0pt]{\textsuperscript{\footnotesize\ensuremath{#1}}}}

\begin{document}
 
\title{Online Container Scheduling for Low-Latency IoT Services in Edge Cluster Upgrade: A Reinforcement Learning Approach
% {\footnotesize \textsuperscript{*}Note: Sub-titles are not captured in Xplore and
% should not be used}
% \thanks{Identify applicable funding agency here. If none, delete this.}
}

\author{
\IEEEauthorblockN{
Hanshuai Cui\IEEEauthorrefmark{1,2},
Zhiqing Tang\IEEEauthorrefmark{1},
Jiong Lou\IEEEauthorrefmark{3,1}, and
Weijia Jia\IEEEauthorrefmark{1,4}}
\IEEEauthorblockA{\IEEEauthorrefmark{1}Institute of Artificial Intelligence and Future Networks, Beijing Normal University, Zhuhai 519087, China}
\IEEEauthorblockA{\IEEEauthorrefmark{2}School of Artificial Intelligence, Beijing Normal University, Beijing 100875, China}
\IEEEauthorblockA{\IEEEauthorrefmark{3}Department of Computer Science and Engineering, Shanghai Jiao Tong University, Shanghai 200240, China}
\IEEEauthorblockA{\IEEEauthorrefmark{4}Guangdong Key Lab of AI and Multi-Modal Data Processing, BNU-HKBU United International College, Zhuhai 519087, China}
\IEEEauthorblockA{hanshuaicui@mail.bnu.edu.cn, zhiqingtang@bnu.edu.cn, lj1994@sjtu.edu.cn, jiawj@bnu.edu.cn}
\thanks{Corresponding authors: Zhiqing Tang and Weijia Jia.}
% \IEEEauthorblockA{Corresponding Author: Michael Shell \quad Email: ms@michaelshell.org}
}

\maketitle

\begin{abstract}
% 容器在边缘计算中轻量级。
% IoT将计算密度的任务卸载到边缘节点。
% 集群升级时环境复杂，需要同时考虑升级状态，资源分布等。
%%% 在本文中，我们提出了一种在集群升级时候，容器的在线调度算法，以减少整体任务延迟。
% 具体来说：1. 对边缘集群中容器的在线调度问题进行了建模，以最小化整体（总延迟）。2.提出了一种基于策略梯度的强化学习算法，做出在线调度决策。3. 在模拟数据上进行了实验，实验结果表明相比于基线算法，我们的算法将总延迟降低了大约20\%。

% 容器轻量化、可扩展性高，非常适合在移动边缘计算中部署。物联网设备将计算密集型任务卸载到边缘节点在容器中运行，减轻了集中式云资源的负担。然而在边缘节点集群升级时候，需要对集群中的容器进行合理调度，以减少对运行中任务的影响。在本文中，我们提出了一种边缘集群升级中的低延迟容器调度算法。具体来说：1. 我们对边缘集群中容器的调度问题进行了建模，以最小化整体（总延迟）。2.提出了一种基于策略梯度的强化学习算法，做出在线调度决策。3. 在模拟数据上进行了实验，实验结果表明相比于基线算法，我们的算法将总延迟降低了大约20\%。

In Mobile Edge Computing (MEC), Internet of Things (IoT) devices offload computationally-intensive tasks to edge nodes, where they are executed within containers, reducing the reliance on centralized cloud infrastructure. Frequent upgrades are essential to maintain the efficient and secure operation of edge clusters. However, traditional cloud cluster upgrade strategies are ill-suited for edge clusters due to their geographically distributed nature and resource limitations. Therefore, it is crucial to properly schedule containers and upgrade edge clusters to minimize the impact on running tasks. In this paper, we propose a low-latency container scheduling algorithm for edge cluster upgrades. Specifically: 1) We formulate the online container scheduling problem for edge cluster upgrade to minimize the total task latency. 2) We propose a policy gradient-based reinforcement learning algorithm to address this problem, considering the unique characteristics of MEC. 3) Experimental results demonstrate that our algorithm reduces total task latency by approximately 27\% compared to baseline algorithms.

\end{abstract}


% --------------------------------------
% TODO: 
% 1. - [x] IoT
% 2. - [x] 算法改名？
% 3. - [x] pod or container?
% 4. - [x] challenges
% 5. - [x] 时态
% 6. - [x] 图片中OPS改成OCS
% 7. - [x] 句式检查一遍
% 8. - [x] Time/Latency 统一
% 9. - [x] node/edge node
% 10. - [x] title
% 11. - [x] 算法/方法
% --------------------------------------

\section{Introduction}
% 1. 第一段介绍一下 edge computing，引出 container、pod、K8S等
% 2. 升级（虚拟机升级，节点升级），发布（蓝绿，滚动），升级为了什么（安全补丁，功能更新），升级造成运行中pod（container）移动，产生延迟。
% 3. 第三段，k8s中pod调度策略（在默认情况下，一个Pod在哪个Node节点上运行，是由Scheduler组件采用相应的算法计算出来的，这个过程是不受人工控制的。）在实际使用中，这并不满足的需求。由于边缘计算中边缘节点的地域分布、计算存储通信资源异构，仅考虑镜像的分布远远不够，位置、资源也要考虑。
% 4. 第四段，与启发对比，RL有更好好的调度策略，可以online decision。基于策略梯度的算法。
% 5. 第五段，总结一下为了解决这个 (Low-Latency Container Scheduling in Edge Cluster Upgrade) 问题，本文基于策略梯度强化学习算法，提出了online container scheduling（OCS）算法。这个算法通过考虑了异构边缘节点的资源，任务的特征，镜像的分布情况，可以做到在线决策，以最大化长期的收益。同时，边缘网络中位置很重要，我们使用了ViT算法提取节点和任务的位置信息。最后，通过模拟实验验证了OCS算法的有效性，并与k8s中现有的pod调度算法进行比较。实验结果表明我们提出算法的性能优于所有的基线算法。
% 6. 总结本文贡献、论文结构
% ---------------------------------------------------------

% 在第五代通信技术(5G)时代，边缘计算(EC)是将计算和数据存储的处理放置在更靠近离用户设备（UE）的网络技术。这种技术可以显著减少UE和数据中心通信所造成的延迟和带宽消耗，更适合处理延迟敏感性任务和服务。随着边缘计算发展，Kubernetes(K8s)、pod和容器等的概念逐渐流行起来\cite{Toka2021,Goethals2022}。Kubernetes \cite{k8s} 是一个用于容器部署（deploying）和编排（orchestration）的开源平台，为部署、管理和扩展容器化应用程序提供了强大的工具。容器轻巧、便携（lightweight, portable），经常被用在边缘计算环境中部署和管理应用程序，方便隔离进程和资源\cite{Ma2019,Wang2021}。pod是K8中最小的可部署单元,由一个或多个容器组成的组（pods are the smallest deployable units in K8S, which can host one or more containers）。

In the era of the Internet of Things (IoT), Mobile Edge Computing (MEC) has emerged as a promising technology that brings computing and data storage closer to IoT devices. This approach significantly reduces latency and bandwidth consumption associated with IoT devices and data center communications, making it more suitable for handling latency-sensitive tasks and services \cite{qian2020}. With the evolution of MEC, containers and Kubernetes are increasingly being used for service deployment \cite{Toka2021,Goethals2022}. Containers are lightweight and portable, frequently employed in MEC to deploy and manage applications while facilitating process and resource isolation \cite{Ma2019,Wang2021,ZhiqingTang2022}. Kubernetes \cite{k8s} is a well-known open-source platform for container orchestration.

%, providing powerful tools for deployment, management, and scaling of containerized applications. 

%Pods are the smallest deployable units in K8s, which can host one or more containers.

% 在边缘计算中，集群的升级可以出于各种原因进行，包括安全补丁、错误修复或添加新功能\cite{guissouma2018empirical,decan2021back}。常见的集群升级方式有：原地升级，蓝绿升级，滚动升级和金丝雀升级等\cite{lwakatare2019devops}。这种升级对于维护边缘集群的安全性和功能至关重要，但缺乏高效的升级策略可能会对用户设备体验产生负面影响。我们重点关注边缘集群升级场景中的低延迟容器调度问题，其中节点的升级可能导致运行的pod（容器）从一个节点迁移到另一个节点，从而导致额外的延迟和资源消耗。因此，升级集群同时尽量减少对运行应用程序的影响是一个具有挑战性的问题。

% issue是问题，challenge是如何解决问题

An edge cluster consists of a network of interconnected edge nodes that collaborate with each other. Cluster upgrades can be performed for various reasons, such as security patches, bug fixes, or the introduction of new features \cite{guissouma2018empirical,decan2021back}. Such upgrades are essential, but inefficient upgrade strategies may negatively impact the IoT device experience. Consequently, minimizing the impact on running tasks during cluster upgrades poses a challenge. Common cluster upgrade strategies include in-place upgrades, blue-green upgrades, rolling upgrades, and canary upgrades \cite{lwakatare2019devops}. However, these strategies are not well-suited for edge clusters due to their excessive resource requirements or inability to accommodate the geographic distribution of edge nodes. Additionally, frequent image pull-downs may result in network congestion and latency. 

%We focus on the low-latency container scheduling problem in the edge cluster upgrade scenario, in which the upgrade of nodes may cause the migration of running containers from one node to another, resulting in additional latency and resource consumption. Therefore, it is a challenge to upgrade the cluster while minimizing the impact on running tasks.

% 在K8S中，Scheduler组件负责按照预定的调度策略将Pod调度到相应的node节点上\cite{}。调度策略根据资源的可用性、用户的偏好和其他约束条件来确定运行pod的最佳节点。然而，由于边缘节点中地域分布，计算、存储、通信资源的异构性以及低延迟的需求，默认调度策略并不总是满足需求。尤其是在集群升级时，仅仅考虑节点的资源是不够的，还必须考虑升级状态、镜像的分布和位置信息等，以制定更有效的调度策略来降低（mitigate）集群升级带来的挑战。

%In K8s, the scheduler component is responsible for scheduling pods to the corresponding node nodes according to a predefined scheduling policy \cite{rossi2020geo,Carrion2023a,rejiba2022custom}. 
% The scheduling policy determines the most suitable nodes based on resource availability, user preferences, and other constraints. However, the default scheduling policy does not always meet the requirements due to the geographical distribution among edge nodes, the heterogeneity of compute, storage, and communication resources, and the demand for low latency. Especially in cluster upgrading, it is not enough to consider only the resources of nodes, but also to consider the upgrade status and image distribution, to develop a more effective scheduling policy to mitigate the effects brought by cluster upgrade.

The upgrade of nodes may cause running containers to be scheduled from one node to another, resulting in additional latency and resource consumption. Thus, another challenge lies in making online scheduling decisions that yield long-term benefits regarding reduced total task latency. Traditional scheduling algorithms primarily involve rule-based, heuristic-based, or optimization-based approaches \cite{alsharif2021rule,mehrabi2018edge,hu2021reconfigurable,chen2021multitask}. Nonetheless, these algorithms cannot optimize long-term minimum latency in dynamic and diverse MEC environments due to limited storage and bandwidth resources. Recently, Reinforcement Learning (RL) algorithms have been widely applied to various optimization problems \cite{sutton2018reinforcement}. The policy gradient-based RL algorithm has demonstrated promising results for optimal resource allocation and scheduling problems in MEC \cite{ZhiqingTang2022}. As such, a policy gradient-based RL algorithm is proposed for making online scheduling decisions.

In this paper, we first model the online container scheduling problem for edge cluster upgrades to minimize the latency of IoT tasks, while accounting for the geographic distribution and limited resources of edge nodes. 
%, assuming that only one node can be upgraded at a time to avoid excessive resource utilization effectively. 
Second, we propose a policy gradient-based Online Container Scheduling (OCS) algorithm. The OCS algorithm considers the heterogeneity of edge nodes, task characteristics, and image distribution to make online scheduling decisions. Finally, we conduct a set of experiments to verify the effectiveness of the OCS algorithm and compare it with existing scheduling algorithms. Experimental results indicate that our proposed algorithm significantly reduces latency and outperforms all baseline algorithms.

In summary, the contributions of this paper are as follows:

\begin{enumerate} 
\item We model the low-latency container scheduling problem in edge cluster upgrade scenarios for the first time to minimize total task latency, including the communication latency, download latency, and computation latency.
\item An OCS algorithm is proposed based on the policy gradient RL that continually makes online scheduling decisions. The algorithm fully considers the distinctive features of MEC, such as geographical distribution and limited computing resources.
\item We conduct simulation experiments to evaluate the effectiveness of the OCS algorithm. Our experimental results demonstrate that our proposed algorithm outperforms all baseline algorithms.
\end{enumerate}

The remainder of the paper is organized as follows. Section II presents the system model and problem formulation. Section III describes the OCS algorithm. Evaluation is discussed in Section IV. Finally, Section V concludes the paper.

\section{System model and problem formulation}
% 在本节中，首先对系统进行了建模，然后定义了问题的cost，最后在线pod调度问题被提出。
In this section, the system is first modeled. Then, the latency is defined. Finally, the OCS problem is formulated.
%% TODO
%% 描述图1

\begin{figure}[t]
    \centering
    \includegraphics[width=0.5\textwidth]{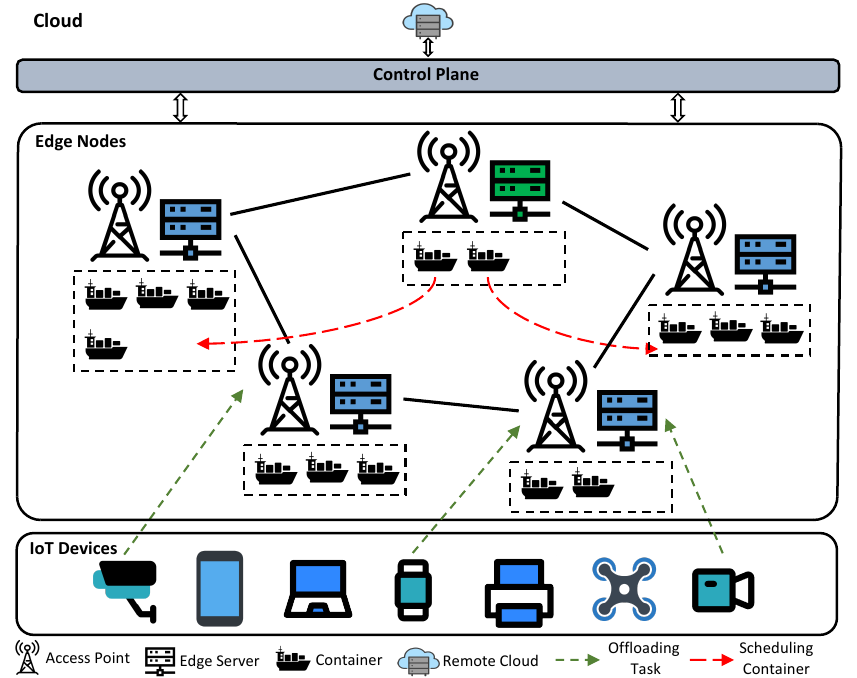}
	\caption{An example of edge cluster upgrade.}
        \label{fig:example}
\end{figure}

\subsection{System Model}
% 我们建模了一个边缘集群的一轮升级场景。User equipments (UEs)中一些计算密集型任务，比如图像识别，文本生成等，需要卸载到边缘节点中执行，然后将结果返回。进一步来说，服务在容器中执行，而且执行前需要请求对应的镜像。由于安全补丁、错误修复或添加新功能等原因，可能会在个时间段进行升级。为了保证服务不中断，在某个节点升级前，需要将节点上所有的pod调度到另一个不在升级中的节点上继运行。同时在升级过程中，不时也会有新任务需要卸载到边缘节点中执行，也需要对它们在哪个节点执行进行决策。集群采用滚动升级策略，集群中所有节点依次进行升级。同时，边缘节点中资（cpu，内存）是有限的，无法调度到不满足资源需求的节点中。除此之外，在集群升级过程中，某个或者某个几个节点会因正在升级导致不可调度。对于不同的任务类型，可能需要不同的容器镜像和资源配额，因此需要设计合理的调度算法和资源管理策略。 For ease of reference, the main notations used in this paper are summarized in TABLE 3.

We model a one-round upgrade scenario for an edge cluster. As illustrated in Fig. \ref{fig:example}, computation-intensive tasks from IoT devices are offloaded to edge nodes, where the results are processed and returned. Tasks are executed in containers, which require the corresponding image to be pulled locally before execution. Upgrades may occur periodically due to security patches, bug fixes, etc. \cite{guissouma2018empirical,decan2021back}. The cluster adopts a rolling upgrade strategy, in which all nodes in the cluster upgrade sequentially, with the node being upgraded shown in green and the node not being upgraded in blue. To ensure uninterrupted service, all containers on a node must be scheduled to another node before upgrading. During the upgrade, new tasks are continuously offloaded to edge nodes, requiring decisions to be made regarding which node they are scheduled on. Meanwhile, resources in edge nodes are limited, and containers cannot be scheduled on nodes that do not meet resource requirements. Additionally, the node being upgraded is set as unschedulable.

The set of nodes is defined as $\mathbf{N} = \{n_1,n_2,\dots,n_{|\mathbf{N}|}\}$, where $|\cdot|$ indicates the number of elements in the set, e.g., $|\mathbf{N}|$ represents the number of nodes. The set of tasks offloaded by different IoT devices to the edge node is $\mathbf {K} = \{k_1,k_2,\dots,k_ {|\mathbf {K}|}\}$. The set of images is denoted as $\mathbf{I} = \{i_1,i_2,\dots,i_{|\mathbf{I}|}\}$, with each image associated with a container. We assume that the resources requested by the task are the same as those occupied by the container, and requesting a container is equivalent to requesting the corresponding image. For ease of reference, the main notations used in this paper are summarized in Table I.

\begin{table}[t]
    \centering
    \caption{Notations}
    \begin{tabular}{ll}
    \bottomrule
    \multicolumn{1}{c}{Notation} & \multicolumn{1}{c}{Definition} \\ \hline
    $\textbf{N}$          & Node set                              \\
    $n$                   & $n^{th}$ node ($n\in\mathbf{N}$)               \\
    $C_n(t)$              & CPU resource of node $n$ at time $t$               \\
    $M_n(t)$              & Memory resource of node $n$ at time $t$               \\
    $D_n(t)$              & Storage capacity of node $n$ at time $t$               \\
    $F_n$                 & CPU frequency of node $n$             \\
    $B_n$                 & Bandwidth of node $n$              \\
    $u_n(t)$              & Upgrade status of node $n$ at time $t$               \\
    % $L_n$                 & Location of node $n$               \\
    % $\textbf{I}$          & Image set             \\
    % $i$                   & $i^{th}$ task ($i\in\mathbf{I}$)               \\
    % $g_i$                 & Size of image $i$             \\
    $\textbf{K}$          & Task set            \\
    $k$                   & $k^{th}$ task ($k\in\mathbf{K}$)               \\
    $c_k$                 & CPU request of task $k$               \\
    $m_k$                 & Memory request of task $k$               \\
    $f_k$                 & CPU frequency request of task $k$               \\
    $q_k$                 & Image request of task $k$               \\
    $d_k$                 & Size of task $k$ \\
    % $t_k$                 & Release time of task $k$               \\
    % $l_k(t)$              & Location of task $k$ at time $t$               \\
    $\mathbf{I}$          & Image set              \\
    $s_i$                 & Size of image $i$               \\
    % $T^{comm}_{n,k}$      & Communication latency for task $k$ on node $n$        \\
    % $T^{down}_{n,k}$      & Download latency for task $k$ on node $n$        \\
    % $T^{comp}_{n,k}$      & Computation latency for task $k$ on node $n$        \\
    % $T^{trans}_{n,k}$     & Total latency for task $k$       \\
    \bottomrule
    \end{tabular}
    \end{table}

\subsection{Latency}

\textbf{Communication latency.}
% 在我们的通信模型中，所有的UE平等的共享节点的带宽。$\eta$ 被定于为为任务k传输到节点n的上行无线传输速率。
In the communication model, all IoT devices equally share the bandwidth of nodes. The uplink wireless transmission rate $\xi_{n,k}$ from task $k$ to node $n$ is defined as \cite{YueWang2019}:
\begin{equation}
    \xi_{n,k} = \frac{B_n}{U_n}log(1+\frac{p_kh_{n,k}}{\sigma^2}),
\end{equation}
% 其中$b_n$表示node的带宽，$U_n$表示同时传输到node的任务数量。$p_k$是任务$k$的传输功率，$h_{n,k}$表示UE与节点之间的传输功率，$\sigma$是高斯白噪声的功率。
where $B_n$ is the bandwidth of node $n$ and $U_n$ is the number of tasks transmitted to node $n$ at the same time. $p_k$ is the transmission power, $h_{n,k}$ is the channel gain between the IoT device and the node, and $\sigma$ represents the power of Gaussian white noise. The communication latency of task $k$ transmitted to node $n$ can be defined as follows:
\begin{equation}
    T^{comm}_{n,k} = \frac{d_k}{\xi_{n,k}}.
\end{equation}

% 此外，与大多数的研究一样，我们忽略了输出的回传通信延迟，因为相比于任务本身来说输出的结果很小。
Furthermore, similar to many studies \cite{Chen2018,Du2018}, we ignore the return communication latency of the result because the result is small compared with the task itself.

\textbf{Download latency.}
% 在k8s中,镜像拉取策略有三种，分别是：IfNotPresent，Always和Never。如果没有指定的话，IfNotPresent是默认策略，它表示如果本地有镜像则使用本地镜像，否则从远程仓库拉取镜像。如果本地有镜像的话，kubelet会跳过镜像拉取操作。换句话说，如果本地有执行任务$k$所需的镜像，则执行任务$k$所需的下载延迟为0。我们定义镜像的下载延迟如下。
%there are three image pulling policies: \textit{IfNotPresent}, \textit{Always}, and \textit{Never}. If not specified, 
% download latency主要是指镜像下载延迟
Download latency refers to image download latency, which is defined as:
\begin{equation}
    T^{down}_{n,k} = x_{n, q_k} \times (\frac{s_{q_k}}{B_n} + T^{queue}_n ),
\end{equation}
% x是用来表示镜像i是否在节点n的二元变量，如果镜像i在节点n上，则x=1，否则，x=0.
where $q_k$ is the image requested by task $k$ and $s_{q_k}$ is the size of the image required to process task $k$. $x_{n,i} \in \{0,1\}$ is the binary variable to indicate whether image $i$ is on node $n$. If $x_{n,i}=1$, image $i$ is on node $n$, otherwise not on node $n$. $T^{queue}_n$ is the queuing download latency on node $n$. Therefore, if the image required to process the task is available locally, the download latency is 0.

\textbf{Computation latency.}
% 不同任务在不同pod中执行，任务之间相互隔离，所有任务并行执行.任务计算延迟可以按以下方式计算。
Different tasks are executed in different containers, and all tasks are executed in parallel. The computation latency can be calculated as follows:
\begin{equation}
    T^{comp}_{n,k} = \frac{f_k}{F_n},
\end{equation}
% 其中$f_k$是任务$k$请求的cpu频率，$F_n$是节点$n$的cpu频率。
where $f_k$ is the CPU frequency requested by task $k$, and $F_n$ is the computing power of node $n$.

% 综上，任务$k$在节点$n$上执行的总延迟可以表示为：
In summary, the total latency of task $k$ execution on node $n$ can be denoted as:
\begin{equation}
    T^{total}_{k} = T^{comm}_{n,k} +T^{down}_{n,k} +T^{comp}_{n,k}.
\end{equation}

\subsection{Problem Formulation and Analysis}

\textbf{Constraints.}
% 1. task资源
% 2. image空间
% 3. 任务位置
% pod需要被分配一定的资源用来执行task，且node的上的资源总量是有限的，如果超过node的资源上限，则有可能导致pod无法正常运行。因此需要对node上pod占用的资源总量进行限制。在k8s中，pod的使用的资源包括cpu和memery。在我们的建模中，默认一个pod中只执行一个task，且pod使用资源量与task请求资源量相同。节点上的资源限制可以表示为：
The containers need to be assigned certain resources, while the total amount of resources on the node is limited. The resource limits on the node can be denoted as:
\begin{equation}
    \sum_{k \in \textbf{K}} y_{n,k} \times c_k \le C_n \ ,\  \sum_{k \in \textbf{K}} y_{n,k} \times m_k \le M_n \ ,\ \forall n,
\end{equation}
% and
% \begin{equation}
%     \sum_{k \in \textbf{K}} y_{n,k} \times m_k \le M_n \ ,\  \forall n
% \end{equation}
% 其中二元变量$y_{n,k}$表示任务k是否在节点n上执行。如果y_{n,k} = 0，则任务k在节点n上执行，否则如果任务k不在节点n上执行。$
where the binary variable $y_{n,k} \in \{0,1\}$ indicates whether task $k$ is executed on node $n$. If $y_{n,k} = 1$, the task $k$ is executed on node $n$. Otherwise, the task $k$ is not executed on node $n$.

% 同时，node上的用于存放image的存储空间是有限的。节点上的存储空间限制可以表示为：
Meanwhile, the storage space for the image on a node is limited, which can be defined as:
\begin{equation}
    \sum_{i \in \textbf{I}} x_{n,i} \times s_i \le D_n \ ,\  \forall n.
\end{equation}

% 此外，与研究[1],[2],[3]一样，任务被认为是不可分的，因此每个task只会被调度到一个node中，可以表示为：
% 这里的参考文献从intro中找几个

Furthermore, tasks are regarded as indivisible, so each task is scheduled to only one node, which can be expressed as:
\begin{equation}
    \sum_{n \in \mathbf{N} } y_k^n=1, \quad \forall k.
\end{equation}

\textbf{Problem Formulation.}
% 我们的目标是从长期角度最大限度地减少总体任务完成时间，这是在Eq中定义的。(3).目标是找到最好的策略，在遵守限制的同时最大限度地减少整体时间。因此，MEC中的LDS问题定义如下：
% 我们目标最小化k8s集群升级的过程中，集群中任务的总体完成时间。The target is to find the best policy to minimize the overall time while obeying the constraints. 因此，在线pod调度算法被定义如下：
We aim to minimize the average total latency of the tasks during the edge cluster upgrade. The target is to find the best policy to minimize the latency while obeying the constraints. The OCS problem is defined as:

\textbf{Problem OCS.}
\begin{equation}
\begin{aligned}
    min &T = \sum_{k \in \textbf{K}}T_k^{total},\\
    s.t. \ & Eqs. \ (6)-(8).
    % & x_{n,i} \in \{0,1\}, \ \forall n \in \textbf{N}, \ \forall i \in \textbf{I}, \\
    % & y_{n,k} \in \{0,1\}, \ \forall n \in \textbf{N}, \ \forall k \in \textbf{K}.
\end{aligned}
\end{equation}

% OCS问题是NP-hard问题，因此传统的算法可能无法在合理的时间内得出最优解。但是使用强化学习可以通过不断学习和优化来逐步得到更好的解决方案。
The OCS problem is NP-hard, so the traditional algorithm may need help to get the optimal solution in a reasonable time. The RL algorithm can gradually lead to a better solution through continuous learning and optimization \cite{wang2020reinforcement}.

% --------------------------------------
% TODO: 
% 1. MDP
% 2. 
% --------------------------------------

% --------------------------------------
% TODO: 
% 1. 算法分析NP-hard
% 2. 
% --------------------------------------

\section{Algorithms}
% 在本节中，首先介绍了算法的设定。然后展示了OCS算法的概述。最后对算法的计算复杂度进行分析。
In this section, the settings of the OCS algorithm are first presented. Then, the OCS algorithm is illustrated. 
%Finally, the computational complexity of the algorithm is analyzed.

\subsection{Algorithm Settings}
% 在本小节中，强化学习中的设定被介绍，包括状态，动作空间和奖励。
In this subsection, the settings in the RL algorithm are introduced, including state, action space, and reward.

\textbf{State.}
% 状态$s_t$中包含几个方面的信息，包括节点信息，任务信息和位置信息。
% 其中，节点信息包括节点的资源信息和节点的升级状态。节点的资源信息包括节点的CPU，内存和存储空间，还有节点的CPU频率和带宽，可以被定义为：
The state $s_t$ contains the node state and task state. 
The node state includes the resource state and the upgrade state. The resource state includes the CPU, memory, and storage capacity of the node at time $t$, as well as the CPU frequency and bandwidth of the node, which can be defined as:
\begin{equation}
\begin{aligned}
    s_t^{node,r} = \{&C_1(t),C_2(t),\dots,C_{|\mathbf{N}|}(t), M_1(t),M_2(t),\dots, \\ 
     &M_{|\mathbf{N}|}(t), D_1(t),D_2(t),\dots,D_{|\mathbf{N}|}(t), \\
    &F_1,F_2,\dots,F_{|\mathbf{N}|}, B_1,B_2,\dots,B_{|\mathbf{N}|}
    \}.
\end{aligned}
\end{equation}

% 节点的升级状态用变量$p_n \in \{0,1,2\}$表示。$p_n = 0$表示节点$n$未升级，$p_n = 1$表示节点$n$正在升级，$p_n = 2$表示节点$n$已经升级完成。所以节点的升级状态可以表示如下：

The upgrade status of the node $n$ at time $t$ is denoted by the variable $p_n(t) \in \{0,1,2\}$. $p_n = 0$ indicates that node $n$ has not been upgraded, $p_n = 1$ indicates that node $n$ is being upgraded, and $p_n = 2$ indicates that node $n$ has been upgraded. Therefore, the upgrade state of nodes can be denoted as:
\begin{equation}
    s_t^{node,u} = \{p_1(t),p_2(t),\dots,p_{|\mathbf{N}|}(t)\}.
\end{equation}

Finally, the state for all nodes is defined as follows:
\begin{equation}
    s_t^{node} = s_t^{node,r} \cup s_t^{node,u}.
\end{equation}

% 任务状态包括任务请求的资源，请求的镜像id，和任务所需的镜像在各个节点上的状态。
The task state includes the status of the images required to execute the task on each node and the requested resources. Thus, the task state can be denoted as follows:
\begin{equation}
\begin{aligned}
    s_t^{task} = \{&x_{1,q_k}, x_{2,q_k}, \dots, x_{|\mathbf{N}|,q_k}, \\
    &t_{1,q_k}, t_{2,q_k}, \dots, t_{|\mathbf{N}|,q_k},c_k, m_k, f_k, d_k, q_k\},
\end{aligned}
\end{equation}
% 其中$t_{n,q_k}$为镜像在每个节点的下载时间，可以通过公式3计算。
where $t_{n,q_k}$ is the download time of the image in each node, which can be calculated by Eq. (3).

%  因此，任务的状态可以表示如下：

% \begin{equation}
%     s_t^{task} = \{s_t^{task,r} \cup s_t^{task,image}\}
% \end{equation}

In summary, the state at time $t$ is defined as:
\begin{equation}
    s_t = s_t^{node} \cup s_t^{task} .
\end{equation}

\textbf{Action space.}
% 在k8s中，用于执行任务的pod默认是由控制层面中的Scheduler组件调度的。OCS算法需要决定pod调度的位置。因此，动作空间就是所有的节点的集合如下：
The container used to execute tasks is scheduled by the scheduler. The OCS algorithm needs to determine which node to schedule. Therefore, the action space is the set of all nodes as follows:
\begin{equation}
    a_t \in \mathbf{A} = \{1,2,\dots,{|\mathbf{N}|}\}.
\end{equation}

\textbf{Reward.} Defining a proper reward is crucial in the RL algorithm. Since different tasks require varying amounts of computation power, considering only the total latency may lead to an unstable training process. Thus, both the expected and actual latencies of the task are included in the reward, which can be defined as follows:
\begin{equation}
    r_t = T^{e}_k - T^{total}_{k},
\end{equation}
where $T^{e}_k = \frac{f_k}{F_m}$ represents the expected total latency of the task, and $F_m$ denotes the minimum value of the node CPU frequency. If the task is completed earlier than expected, the reward is positive, with the completion time being inversely proportional to the reward. Conversely, the reward is smaller. From a long-term perspective, the cumulative reward is $R_t = \sum^{T}_{t=0}\gamma^t r_t$, where $\gamma$ is the discount factor with a value ranging between [0, 1].

\subsection{Online Container Scheduling}

\begin{figure}[t]
    \centering
    \includegraphics[width=0.45\textwidth]{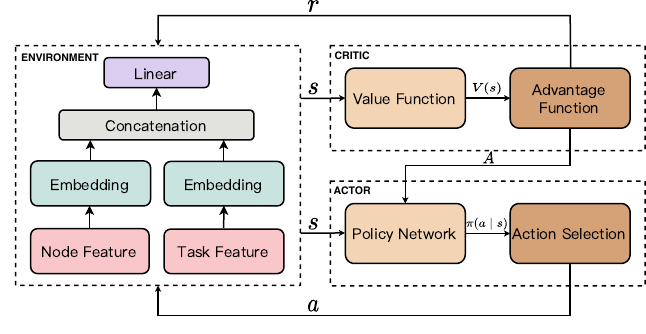}
	\caption{Overview of the OCS algorithm.}
    \end{figure}

\textbf{Overview.} The OCS algorithm considers the heterogeneity of edge nodes and the characteristics of tasks in the edge cluster. The framework of the OCS algorithm is depicted in Fig. 2. Specifically, the node state and task state can be observed from the environment. After obtaining their features, they are embedded and concatenated, and then fed into the policy network to make the corresponding scheduling operations. The reward is subsequently obtained based on the action. Finally, the policy network and value functions are updated using the policy gradient-based algorithm.

\textbf{Training.} The OCS algorithm is based on policy optimization. Policy gradient \cite{Sutton1999} is an RL algorithm that directly optimizes the expected return policy. Let $\pi_{\theta}$ denote a policy with parameters $\theta$. The Proximal Policy Optimization (PPO) algorithm \cite{Schulman2017} is based on the policy gradient algorithm, which ensures effectiveness with low computational complexity.
The optimization objective of PPO is as follows:
\begin{equation}
    \begin{aligned}
        \theta_{k+1}=\arg \max _\theta \mathcal{L}^{PPO}\left(\theta_k, \theta\right),
    \end{aligned}
    \end{equation}
and
\begin{equation}
    \begin{aligned}
    \mathcal{L}^{PPO}(\theta_k, \theta)=&\underset{s, a \sim \pi_{\theta_k}}{\mathrm{E}}[(\frac{\pi_\theta(a \mid s)}{\pi_{\theta_k}(a \mid s)} A^{\pi_{\theta_k}}(s, a), \\ &\operatorname{clip}(\frac{\pi_\theta(a \mid s)}{\pi_{\theta_k}(a \mid s)}, 1-\epsilon, 1+\epsilon) A^{\pi_{\theta_k}}(s, a))],
\end{aligned}
\end{equation}
% 优势函数$A^{\pi_\theta}\left(s_t, a_t\right)的定义为：clip()为截断函数，目的是把x限制在[l,r]范围内。\epsilon 是一个超参数，表示进行截断（clip）的范围。
where $clip(x,y,z) = max(min(x,z),y)$ is a clip function to limit $x$ to the range of $[y,z]$ and $\epsilon$ is a hyperparameter that represents the range of clips. Besides, PPO adopts the Generalized Advantage Estimator (GAE) \cite{Schulman2018} to compute the advantages, which can be calculated by:
\begin{equation}
\begin{aligned}
\hat{A}_t=\delta_t+(\gamma \lambda) \delta_{t+1}+\cdots+\cdots+(\gamma \lambda)^{T-t+1} \delta_{T-1},
\end{aligned}
\end{equation}
where $\lambda$ is the GAE parameter, $\delta_t=r_t+\gamma V\left(s_{t+1}\right)-V\left(s_t\right)$ is the TD-error at time step $t$, and $V$ is an approximate value function. 

\begin{algorithm}[t]
    \caption{The OCS Algorithm}
    \SetAlgoLined
    \KwIn{Initial policy parameters $\theta$, initial value function parameters $\phi$, clipping threshold $\epsilon$}%输入参数
    \KwOut{$a_t$}%输出
    % \KwResult{Write here the result}
    
    \For{episode $\leftarrow$ 0,1,2,\dots}{
        Initialize replay memory $\mathbf{D} = \emptyset$ \;
        \For{time slot t $\leftarrow$ 0,1,2,\dots}{
            
            Get the current state $s_t$ \;
            Select action $a_t$ according to $\pi_{\theta}(a_t \mid s_t)$ \;
            Execute action $a_t$ and obtain the reward $r_t$ \;
            Get the next state $s_{t+1}$ \;
            Store transition $(s_t, a_t, r_t, s_{t+1})$ in $\mathbf{D}$\;
        }
        % Compute advantage estimates $\hat{A}$ and store transition $(s_t, a_t, r_t, s_{t+1})$ in $\textbf{D}$
        \For{training step k $\leftarrow$ 0,1,2,\dots}{
            Estimate advantages $\hat{A}_k$ by Eq. (19) \;
            Compute the policy update by Eq. (18) \;
            Update the policy by maximizing the objective function in Eq. (17)\;
        }
    }

    \end{algorithm}

% Compared with the standard policy gradient algorithm that updates the gradient only once per sample, TRPO and PPO can use the same sample for multiple gradient updates. Therefore, TRPO and PPO achieve the same performance as the policy gradient method with less experience. Meanwhile, PPO does so by clipping gradient if the updated policy is not close to the policy used to sample the data.

% 在每个episode中，重放内存$\mathbf{D}$首先被初始化。如3-9行所示, 对于每个time slot $t$，首先获取到当前时刻t的环境观测状态$s_t$，然后根据当前的策略选择动作$a_t$并计算出reward$r_t$,接着获取下一时刻的状态$s_{t+1}$ 最后将transition存储到replay memory中.如10-14行所示, 对于每个training step $k$, 首先根据收集到的一组轨迹计算优势估计，然后采用Adam 随机梯度上升算法最大化 PPO-Clip 的目标函数来更新策略。最后所有的episode结束后，输出结果。

% TODO
% 修改
The OCS algorithm is presented in Algorithm 1. The replay memory $\mathbf{D}$ is first initialized for each episode. As shown in Lines 3 - 9, for each time slot $t$, the observation state $s_t$ of the current time slot $t$ is first obtained, then the action $a_t$ is selected according to the policy, and the reward $r_t$ is calculated. Afterward, the next state $s_{t+1}$ is obtained. Finally, the transition is stored in the replay memory $\mathbf{D}$. As shown in Lines 10 - 14, for each training step $k$, the advantage estimation $\hat{A}_k$ is first computed based on the collected set of trajectories. Then, the stochastic gradient ascent algorithm with Adam is used to maximize the objective function to update the policy. Finally, the results are output after all episodes are completed.

% --------------------------------------   
% TODO: 
% 1. 复杂度分析
% 2. 
% --------------------------------------
\begin{figure*}[t]
    \centering
    \begin{subfigure}[t]{0.245\textwidth}
           \centering
           \includegraphics[width=\textwidth]{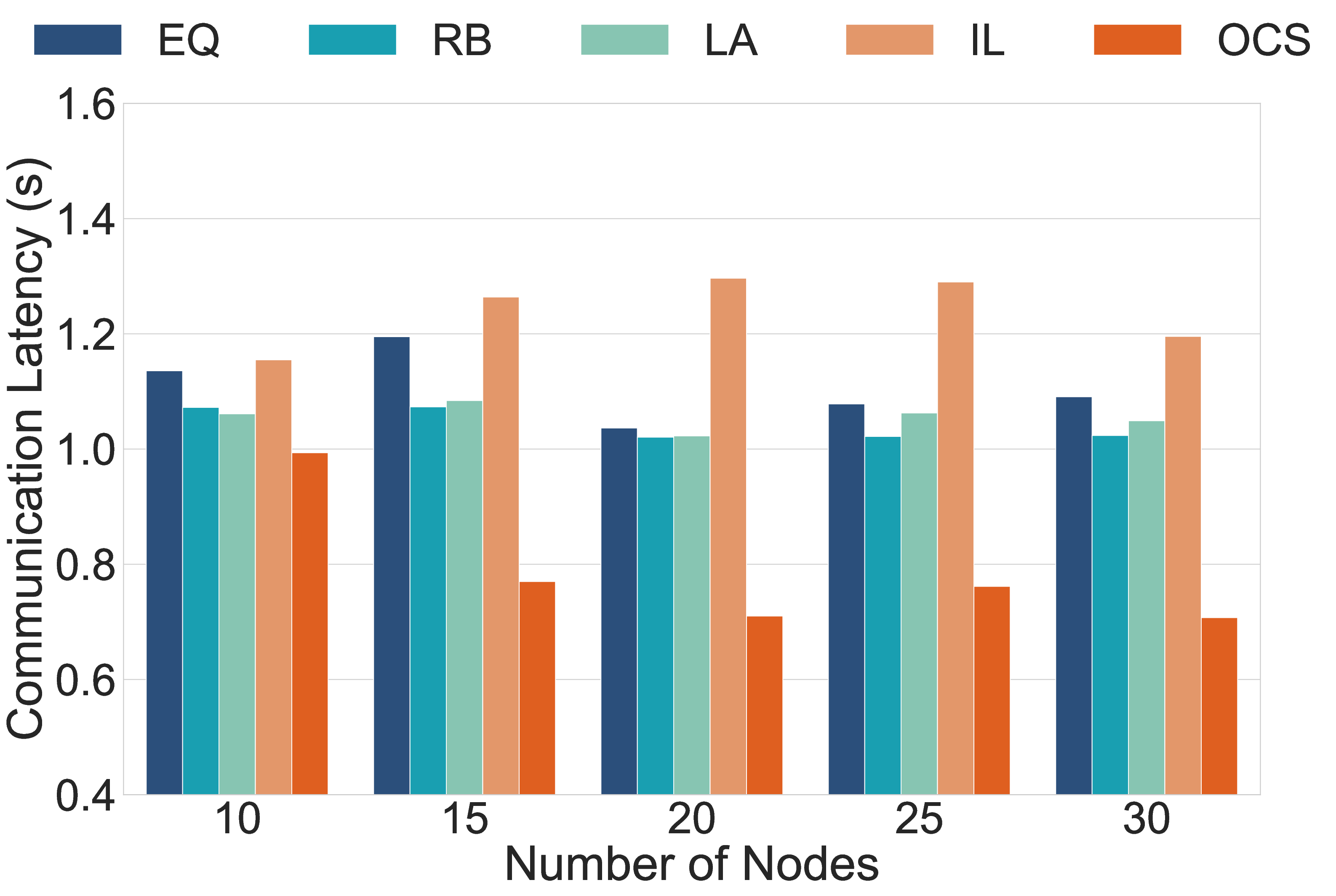}
            \caption{Communication latency}
    \end{subfigure}
    \begin{subfigure}[t]{0.245\textwidth}
            \centering
            \includegraphics[width=\textwidth]{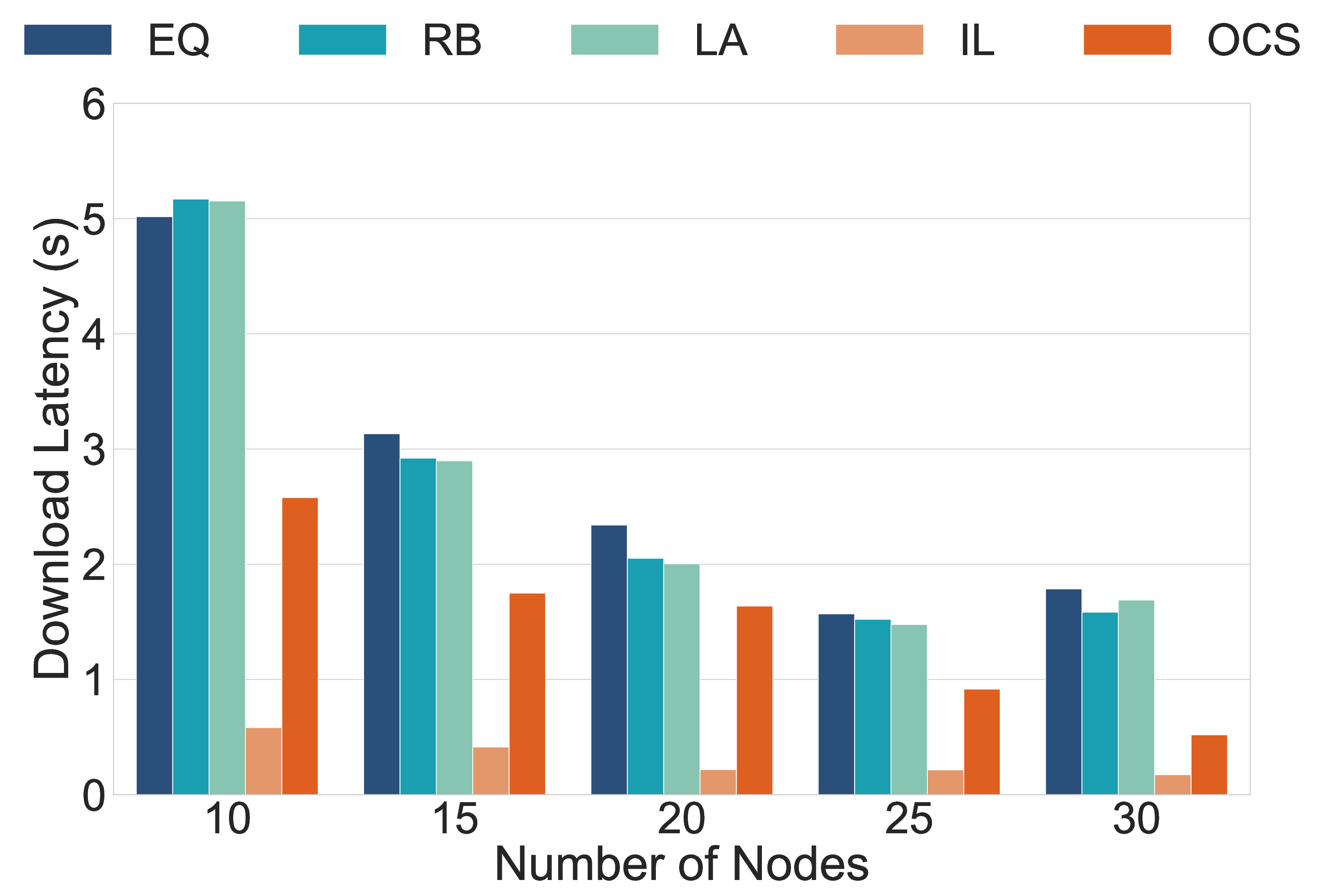}
            \caption{Download latency}
    \end{subfigure}
    \begin{subfigure}[t]{0.245\textwidth}
            \centering
            \includegraphics[width=\textwidth]{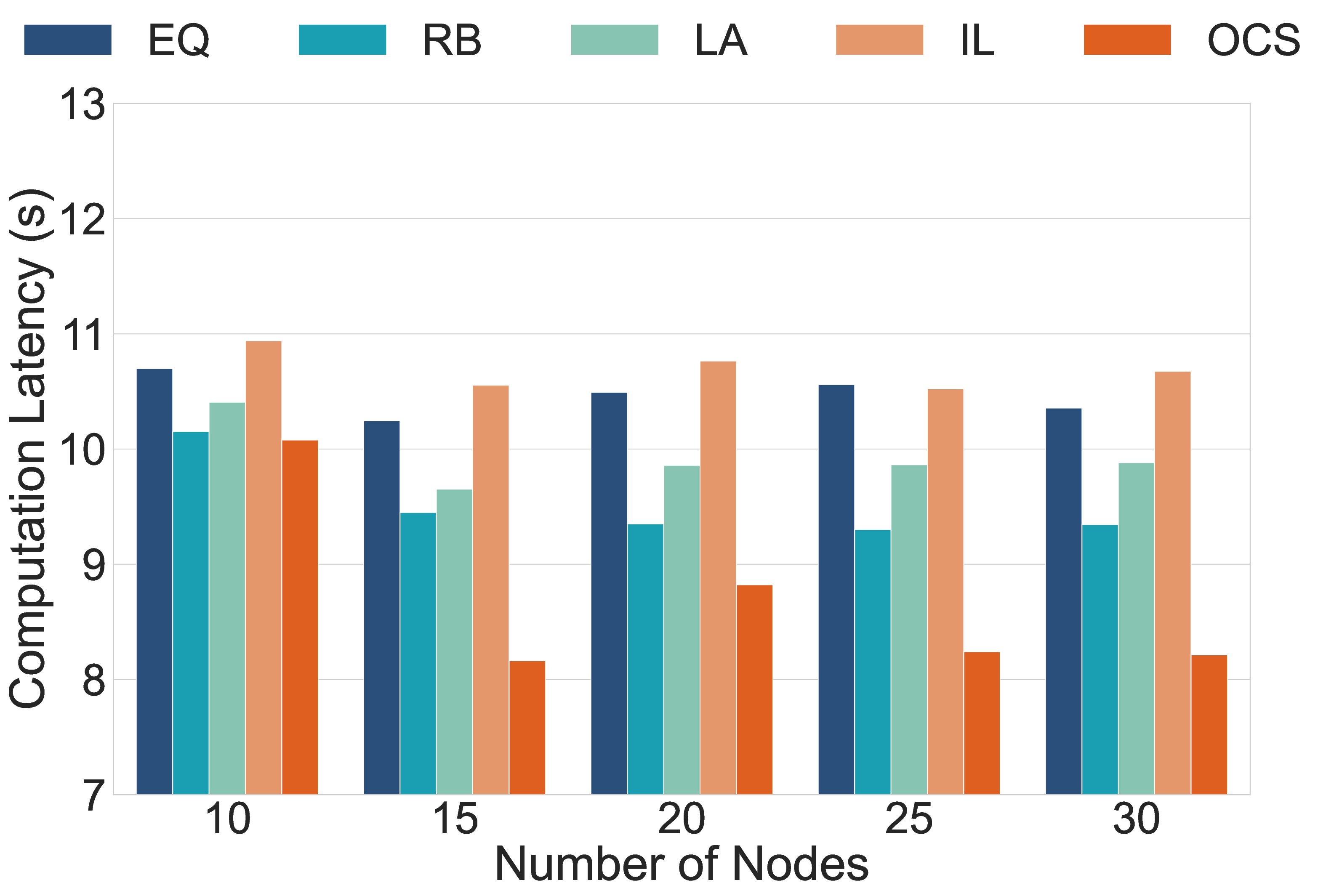}
            \caption{Computation latency}
    \end{subfigure}
    \begin{subfigure}[t]{0.245\textwidth}
        \centering
        \includegraphics[width=\textwidth]{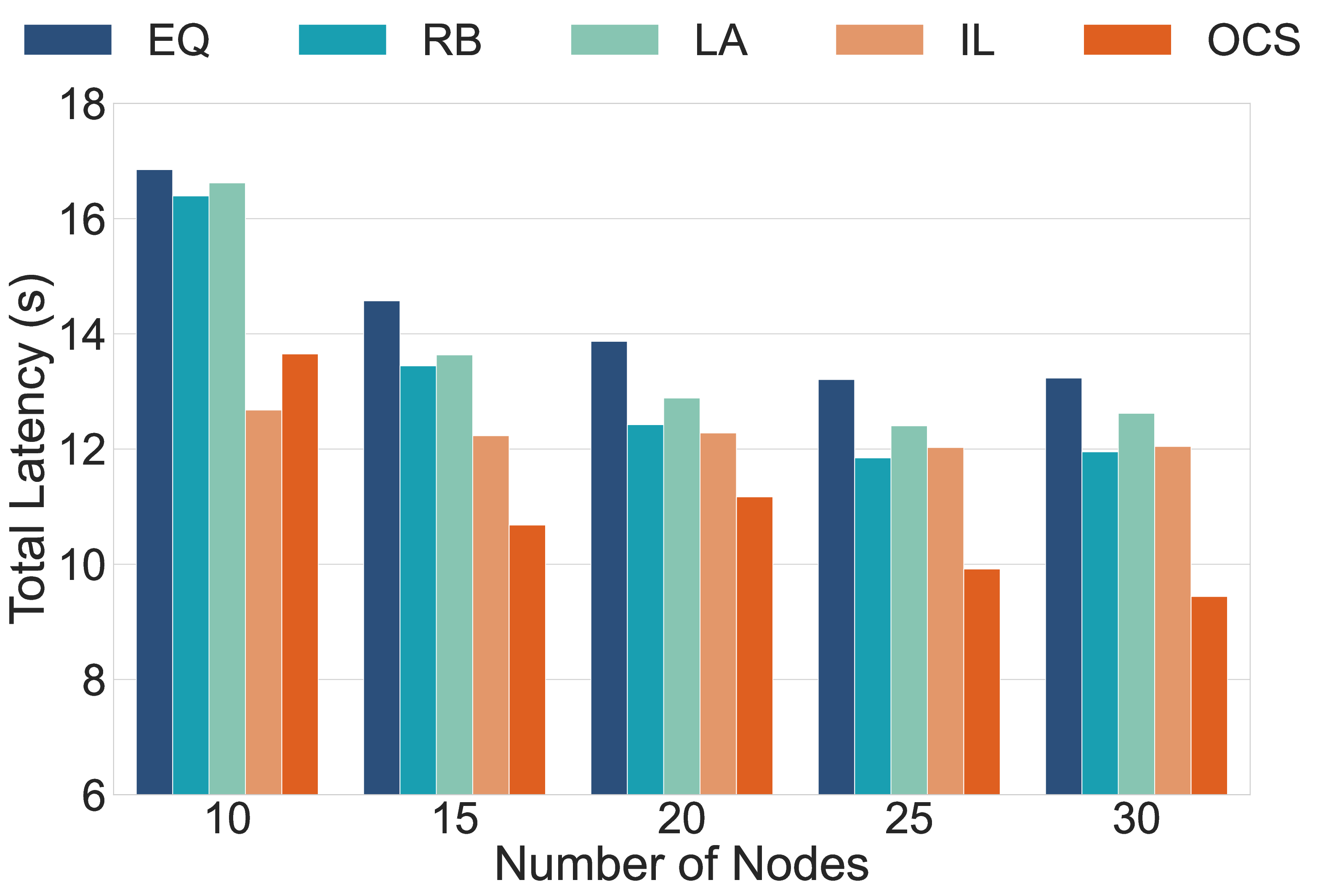}
        \caption{Total latency}
\end{subfigure}
    \caption{Performance with different number of nodes}
\end{figure*}

\begin{figure*}[htpb]
    \centering
     \begin{subfigure}[t]{0.245\textwidth}
            \centering
            \includegraphics[width=\textwidth]{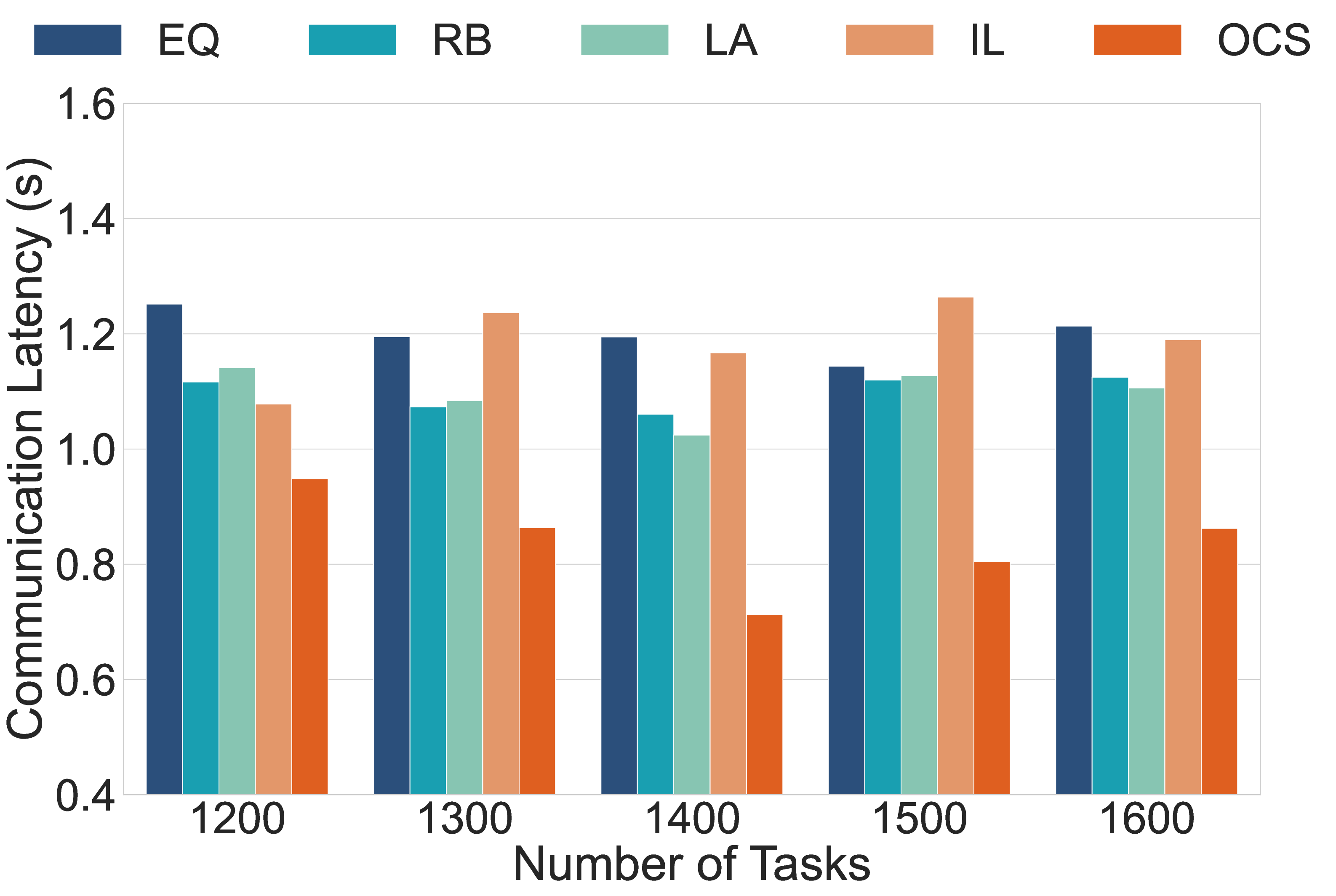}
             \caption{Communication latency}
     \end{subfigure}
     \begin{subfigure}[t]{0.245\textwidth}
             \centering
             \includegraphics[width=\textwidth]{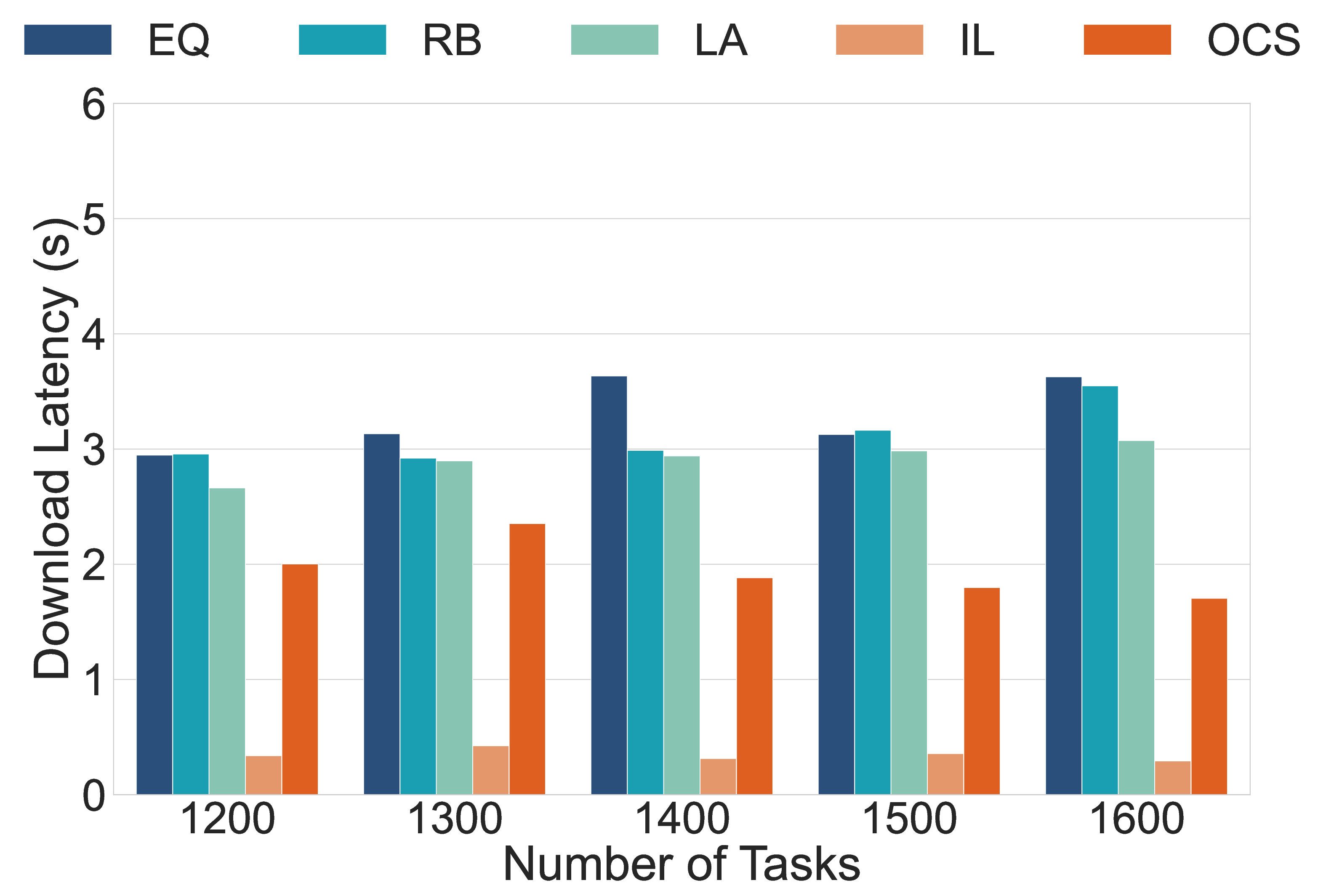}
             \caption{Download latency}
     \end{subfigure}
     \begin{subfigure}[t]{0.245\textwidth}
             \centering
             \includegraphics[width=\textwidth]{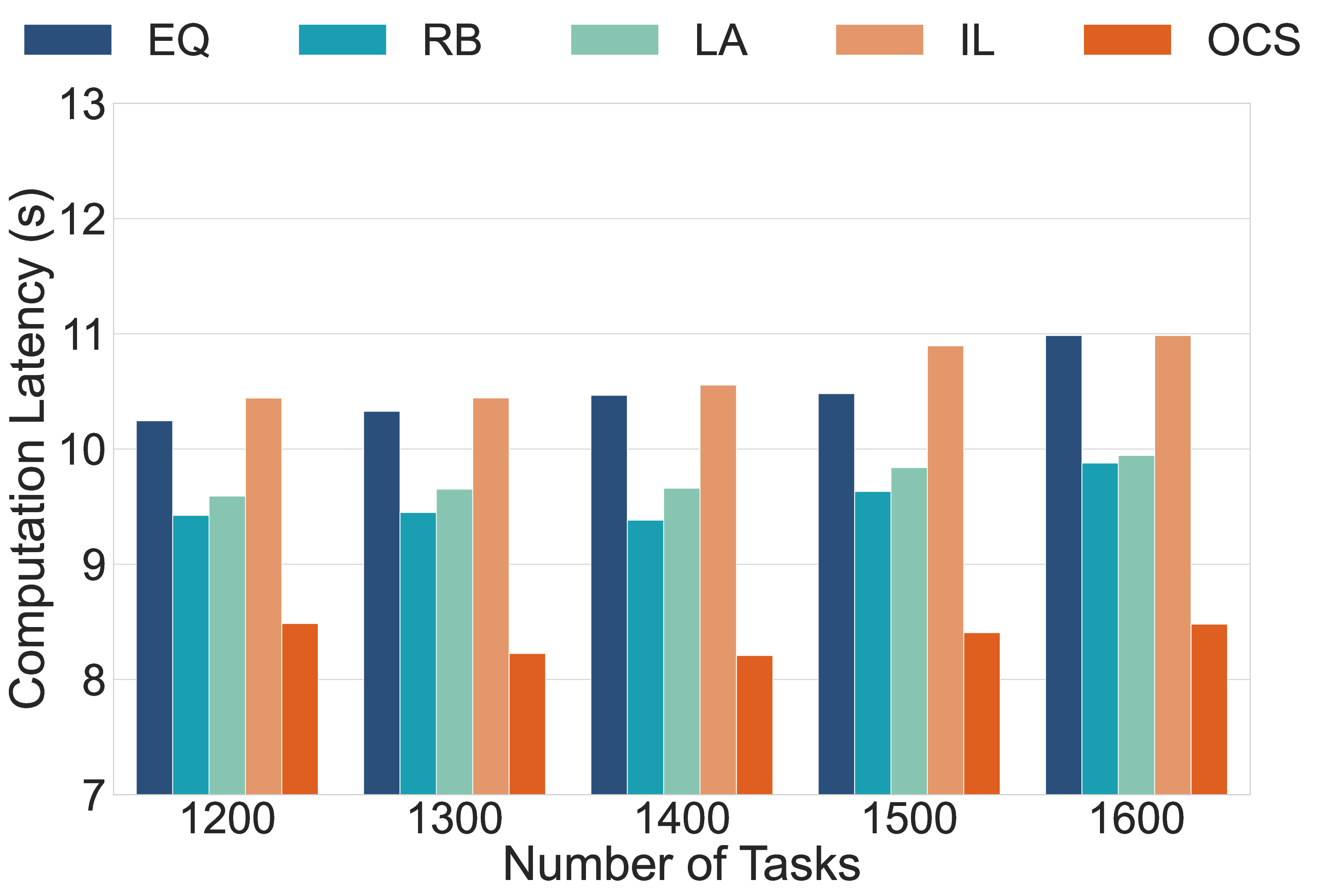}
             \caption{Computation latency}
     \end{subfigure}
     \begin{subfigure}[t]{0.245\textwidth}
         \centering
         \includegraphics[width=\textwidth]{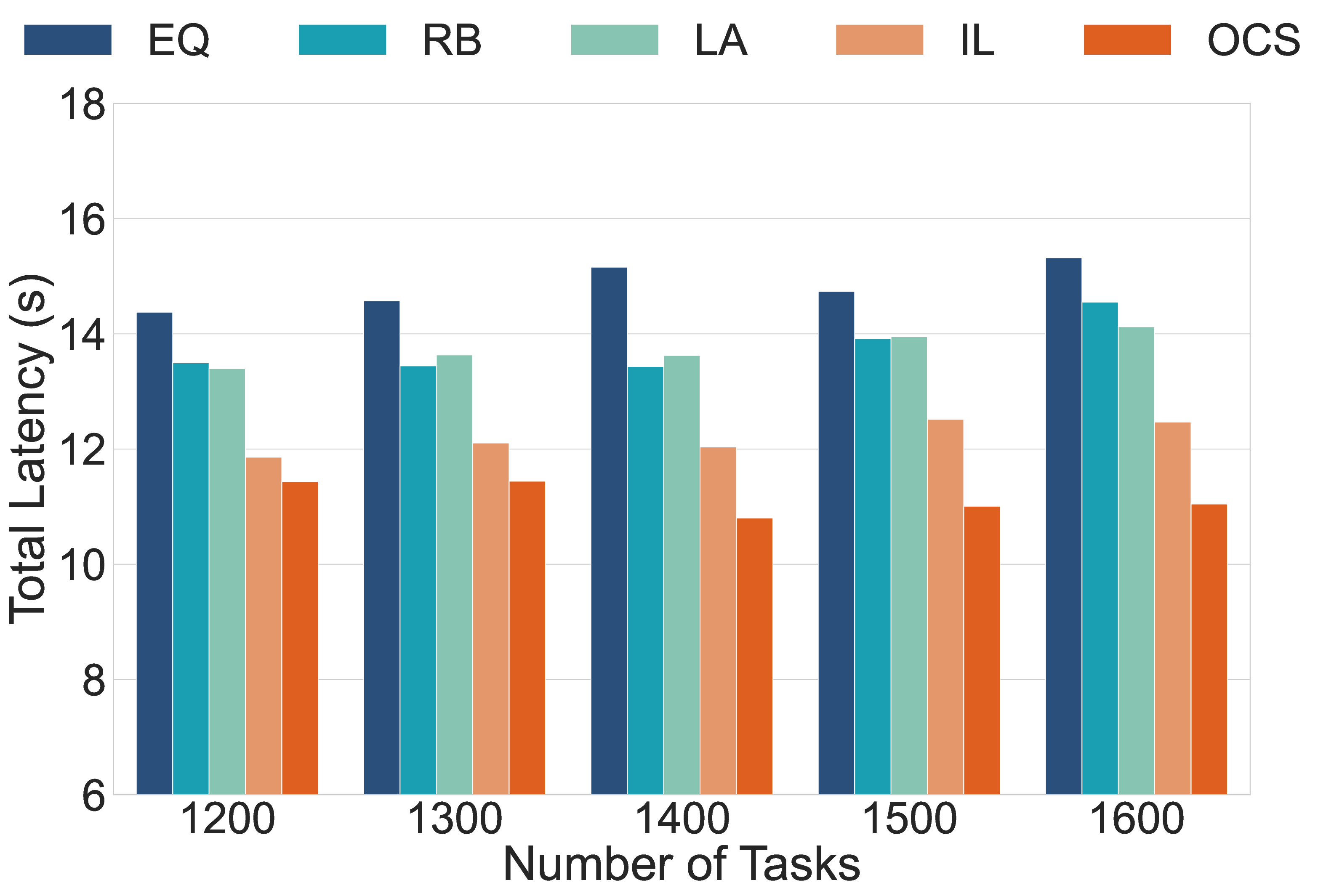}
         \caption{Total latency}
 \end{subfigure}
     \caption{Performance with different number of tasks}
 \end{figure*}

\begin{figure*}[htpb]
    \centering
    \begin{subfigure}[t]{0.25\textwidth}
           \centering
           \includegraphics[width=\textwidth]{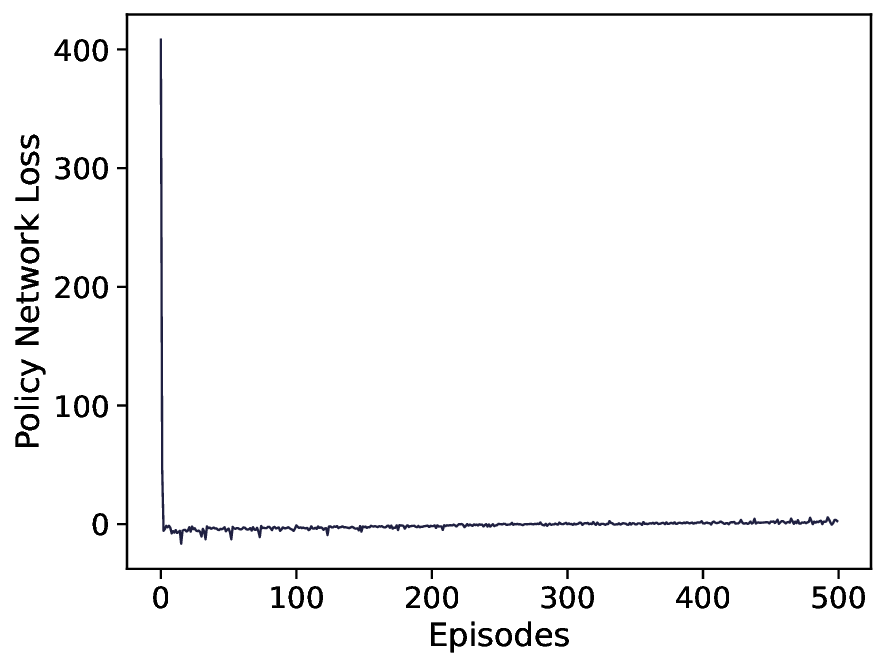}
           \caption{Policy network loss}
    \end{subfigure}
    \begin{subfigure}[t]{0.2455\textwidth}
            \centering
            \includegraphics[width=\textwidth]{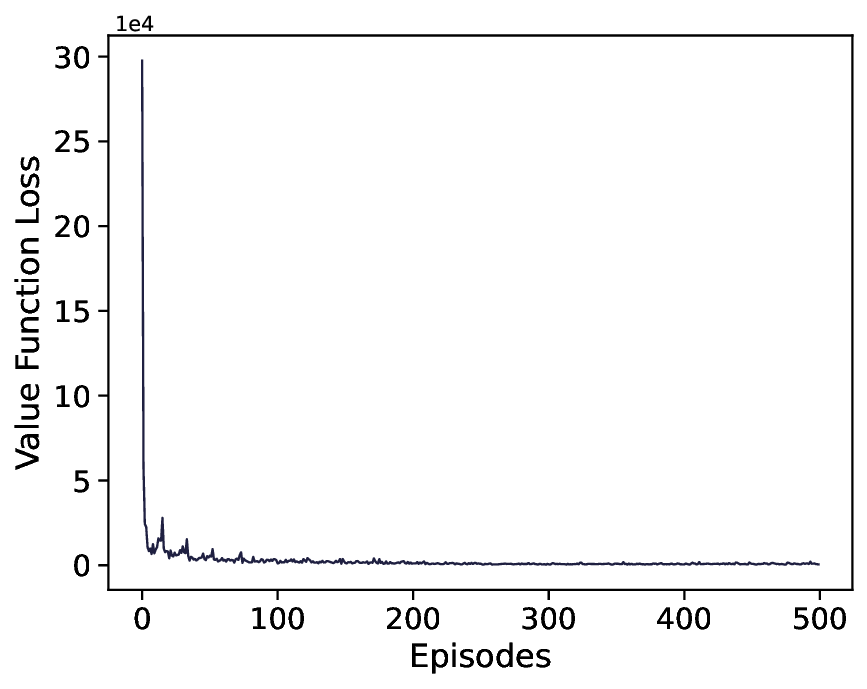}
            \caption{Value function loss}
    \end{subfigure}
    \begin{subfigure}[t]{0.252\textwidth}
            \centering
            \includegraphics[width=\textwidth]{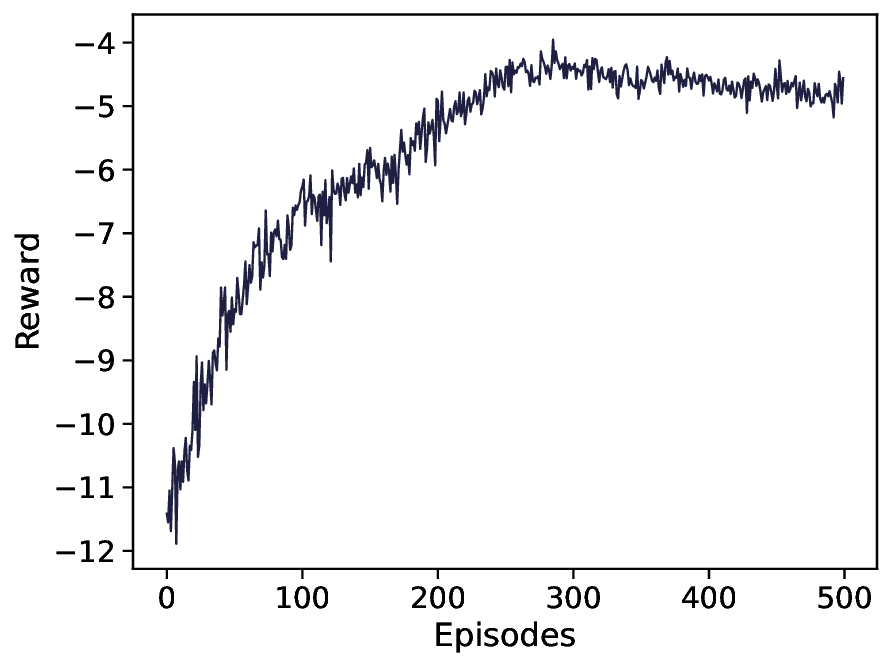}
            \caption{Reward}
    \end{subfigure}
    \caption{Policy network Loss, value function Loss, and reward of the OCS algorithm}
\end{figure*}

\section{Evaluation}
% 在本节中，我们将深入探讨OCS算法的评估，通过模拟实验的方式进行。首先，实验的环境和参数的设置被给出。然后我们比较OCS算法的结果与一些基线算法的结果，来评估OCS算法的性能和效果。最后我们可以通过可视化图表呈现实验结果并对实验结果进行分析和讨论。
In this section, we will delve into the performance of the OCS algorithm through simulation experiments. 
%First, the environment and parameters of the experiment are given. Then we compare the results of the OCS algorithm with some baseline algorithms to evaluate the performance and effectiveness of the OCS algorithm. Finally, we present the experimental results through visual charts with analysis and discussion. 

\subsection{Experimental Settings}
\textbf{Parameter settings.}
% 根据\cite{XuChen2015,YueWang2019}，我们设置传输功率$p=23dBm$, the noise power spectrum density $\sigma = -174(dBm/Hz)$. According to the physical interference model \cite{Funk}, the channel gain between the UE and the node $h_{n,k} = d^{-\alpha}_{n,k}$, 其中$d_{n,k}$是UE和节点之间的距离and $\alpha = 4$ is the path loss factor. The communication bandwidth between the UE and the node is set to [100, 200] MB/s.
Similar to \cite{XuChen2015,YueWang2019}, we set the transmission power $p=23dBm$ and the noise power spectrum density $\sigma=-174 dBm/Hz$. According to the physical interference model \cite{Funk}, the channel gain between the IoT device and the node $h_{n,k}$ is set to $d^{-\alpha}_{n,k}$, where $d_{n,k}$ is the distance between the IoT device and the node and $\alpha = 4$ is the path loss factor. The communication bandwidth between the IoT device and the node is set to [100, 200] Mb/s.

% 模拟区域的大小为$L m \times W m$, and $L$ and $W$ are the length and width of the simulated area, respectively.模拟区域的大小会随着节点的数量增多而增大，默认的大小为$100 m \times 100 m$。 所有的节点都是异构 且随机分布在其中，默认的节点数量是10个。节点的cpu容量设置在[80,120]核之间，cpu频率设置在[15,35]GHz之间，节点的内存设置在[70,130]Gb之间。任务是在每一个time slot随机生成在模拟区域的任意位置，and the task sizes are set from 10Kb to 10Mb。每个任务会请求一种镜像，请求镜像的种类服从正态分布。 如果task被调度到的节点上不存在此镜像，则需要从远程仓库进行下下载，镜像的大小设置from 300Mb to 1.5Gb.在最开始，每个节点上都会执行一定数量任务，并且任务对应的镜像也会存在节点上。 对于神经网络的输入，首先需要将他们缩放到同一维度中。 The hyperparameters of OCS algorithm are listed in Table II.

The area of the simulation region increases as the number of nodes increases, and the default area is $100 m \times 100 m$. All nodes are heterogeneous and randomly distributed, and the default number of nodes is 15. The CPU capacity of the node is set between [80,120] cores. The CPU frequency is set between [15,35] GHz, and the memory is set between [70,130] GB. The task is randomly generated in the simulation region, and the task sizes are set from 10 KB to 10 MB. The types of requested images follow the normal distribution. 
% If the image does not exist on the node to which the task is scheduled, it needs to be downloaded from a remote repository, and the size of the image is set from 300 MB to 1.5 GB. 
%At the beginning, a certain number of tasks are executed on each node and the images corresponding to the tasks are also existed on the nodes. 
% The input of the neural network is scaled to the same order of magnitude. 
The hyperparameters of the OCS algorithm are listed in TABLE II.

\begin{table}[t]
    \caption{Hyperparameter Settings}
    \centering
    \begin{tabular}{lll}
    \bottomrule
    \multicolumn{1}{c}{Type} & \multicolumn{1}{c}{Hyperparameter} & Value                \\ \hline
    Actor                    & Hidden layers                      & 2 Full connection (128,64) \\
                             & Learning rate                      & 1e-4                 \\
    Critic                   & Hidden layers                      & 2 Full connection (128,64) \\
                             & Learning rate                      & 3e-4                 \\
                            %  & Loss Function                      & MSELoss              \\
    Other                    & Discount factor $\gamma$           & 0.98    \\
                             & GAE parameter $\lambda$            & 0.95    \\
                             & Clipping threshold  $\epsilon$     & 0.2     \\
                             & Batch size                         & 32                   \\
                            %  & Activation function                & ReLU                 \\
                             & Optimizer                          & Adam                 \\ 
    \bottomrule
    \end{tabular}
    \end{table}

\textbf{Baselines.} 
% 我们将提出的OCS算法与几个基线算法进行比较，以证明我们提出的算法的有效性。其中(1)-(4)为k8s中几个内置的pod自动调度策略，(5)为基于策略梯度的强化学习算法，(6)为我们提出的算法。基线算法的具体细节如下：
We compare the OCS algorithm with several baseline algorithms to demonstrate the effectiveness of our proposed algorithm: (1) \textbf{EQ}. EqualPriority (EQ) sets the weight of all nodes to 1. (2) \textbf{RB}. ResourcesBalanced (RB) prioritizes balancing the resource usage of each node. (3) \textbf{LA}. LeastAllocated (LA) is a scheduling policy related to the resource usage of the node. (4) \textbf{IL}. ImageLocality (IL) considers the local existence of the image requested by the task. These baselines are built-in scheduling policies in Kubernetes. Moreover, LA and IL are greedy algorithms that select nodes with more resources or images.

% \textbf{Simulator setup.}
% % 我们所有的实验都是在GPU服务器上模拟进行的。我们使用了编程语言python和开源深度学习库（library）pytorch,主要实现了节点，任务，镜像等几个phthon类。Moreover, 我们提出的OCS是一种在线算法，也就是说算法无法预知未来的环境状态，网络更新，动作选择都只与当前状态有关。
% All our experiments are simulated on a GPU server. We use the programming language Python and the open source deep learning library Pytorch and mainly implement several python classes such as nodes, tasks, and images. Moreover, the OCS is an online algorithm, which means that the algorithm cannot foresee the future state of the environment. Network updates and action selection are only related to the current state.

\subsection{Experimental Results}
% 在本小节中，我们对OCS算法在k8s集群升级时pod调度的表现进行了研究。为了验证OCS算法的有效性，我们进行了多个实验，旨在比较OCS算法与其他几个基线算法在不同的设定（如节点数量、任务数量等）下的任务延迟表现。接着，我们对OCS算法的训练过程进行了详细的展示。
% In this section, we examine the container scheduling performance of the OCS algorithm for edge cluster upgrade. 
%To verify the effectiveness of the OCS algorithm, we have conducted many experiments to compare the task latency performance of the OCS algorithm and several other baseline algorithms under different settings (e.g., the number of nodes, the number of tasks, etc.). Then, we demonstrate the training process of the OCS algorithm in detail.

\textbf{Performance with different numbers of nodes.} Fig. 3 shows the average total latency as the number of nodes increases, including communication latency, download latency, computation latency, and total latency. As seen from this figure, the average latency of tasks decreases as the number of nodes increases. The reason is that more nodes are available for scheduling as the number of nodes increases. The scheduler can schedule the containers to more suitable nodes, such as those with a closer distance or more resources. As a result, the average total latency of the task becomes smaller. On the whole, the total latency relationship is OCS $<$ IL $<$ RB $<$ LA $<$ EQ. Therefore, the OCS algorithm performs the best regardless of the number of nodes. Specifically, the total latency with different numbers of nodes is reduced by 40\%, 33\%, 27\%, and 26\% on average compared with EQ, LA, RB, and IL algorithms, respectively.

\textbf{Performance with different numbers of tasks.} The variation of average task latency as the number of tasks increases is illustrated in Fig. 4. The results indicate that the OCS algorithm performs best. While the IL algorithm performs slightly less, the RB and LA algorithms are very close, and the EQ algorithm performs the worst. Overall, as the number of tasks increases, the relationship between the performance of different algorithms in total latency is OCS $<$ IL $<$ LA $<$ RB $<$ EQ. Compared to the EQ, RB, LA, and IL algorithms, the total scheduling latency for the OCS algorithm is reduced by 38\%, 31\%, 27\%, and 12\%, respectively.

\textbf{Performance of the OCS algorithm.} Fig. 5 shows the convergence of the OCS algorithm. The policy network loss and value function loss of the OCS algorithm have large values at the beginning of training. However, as the training steps increase, both decrease rapidly and eventually fluctuate near a specific value, indicating that the algorithm has converged.

\section{Conclusion}
% 提出了集群升级时候，容器的低延迟在线调度算法。
% 首先，对在线容器调度问题进行了建模。
% 其次，使用基于策略梯度的强化学习算法来进行在线决策。
% 最后，实验结果表明我们的算法在总延迟上比基线算法低20%。
% 未来会在K8s中实现此方法。
% 本文中提出了一种边缘集群升级中的低延迟容器调度算法。首先，我们对在线容器调度问题进行了建模。其次，使用基于策略梯度的强化学习算法来进行在线决策。最后，实验结果表明我们的算法在总延迟上比基线算法低20%。未来我们会在K8s中实际部署此方法。

This paper proposes a low-latency container scheduling algorithm for IoT services in edge cluster upgrades. First, we comprehensively model the OCS problem, considering communication, download, and computation latency. Second, a policy gradient-based RL algorithm is proposed to make online scheduling decisions, which fully considers the distinctive features of MEC. Finally, experiments are conducted on a simulated edge cluster, and the experimental results demonstrate that our algorithm achieves approximately 27\% lower total latency compared to the baseline algorithm. In future work, we will deploy this algorithm in the Kubernetes system.

\section*{Acknowledgment}
This work is supported in part by the Guangdong Key Lab of AI and Multi-modal Data Processing, United International College (UIC), Zhuhai under Grant 2020KSYS007 sponsored by Guangdong Provincial Department of Education; in part by the Chinese National Research Fund (NSFC) under Grant 62272050; in part by Institute of Artificial Intelligence and Future Networks (BNU-Zhuhai) and Engineering Center of AI and Future Education, Guangdong Provincial Department of Science and Technology, China; Zhuhai Science-Tech Innovation Bureau under Grants ZH22017001210119PWC and 28712217900001, and in part by the Interdisciplinary Intelligence SuperComputer Center of Beijing Normal University (Zhuhai).

\bibliographystyle{IEEEtran}
\bibliography{ref}

\end{document}